\documentclass{ws-ijmpa}
\usepackage[super,compress]{cite}
\usepackage{graphicx}
\usepackage[dvipsnames]{xcolor}
\usepackage{url}
\usepackage[]{hyperref}
\begin{document}
\markboth{Anna Feh\'erkuti}{XY Factorization Bias in Luminosity Measurements}

%
\catchline{}{}{}{}{}
%

\title{XY Factorization Bias in Luminosity Measurements}

\author{Anna Feh\'erkuti\footnote{HUN-REN Wigner Research Centre for Physics, Konkoly-Thege Mikl\'os road 29-33, Budapest, 1121, Hungary}}

\address{Department of Atomic Physics, E\"otv\"os Lor\'and University, P\'azm\'ny P\'eter s\'et\'any 1/A\\
Budapest, 1117, Hungary\\
feherkuti.anna@wigner.hun-ren.hu, anna.feherkuti@ttk.elte.hu}

\author{P\'eter Major}

\address{Department of Atomic Physics, E\"otv\"os Lor\'and University, P\'azm\'ny P\'eter s\'et\'any 1/A\\
Budapest, 1117, Hungary\\
peter.major@ttk.elte.hu}

\author{Gabriella P\'asztor}

\address{Department of Atomic Physics, E\"otv\"os Lor\'and University, P\'azm\'ny P\'eter s\'et\'any 1/A\\
Budapest, 1117, Hungary\\
gabriella.pasztor@ttk.elte.hu}

\maketitle

\begin{history}
\received{Day Month Year}
\revised{Day Month Year}
\end{history}

\begin{abstract} 
For most high-precision experiments in particle physics, it is essential to know the luminosity at highest accuracy. The luminosity is determined by the convolution of particle densities of the colliding beams. In special van der Meer transverse beam separation scans, the convolution function is sampled along the horizontal and vertical axes with the purpose of determining the beam convolution and getting an absolute luminosity calibration. For this purpose, the van der Meer data of luminometer rates are separately fitted in the two directions with analytic functions giving the best description. With the assumption that the 2D convolution shape is factorizable, one can calculate it from the two 1D fits. The task of XY factorization analyses is to check this assumption and give a quantitative measure of the effect of non-factorizability on the calibration constant to improve the accuracy of luminosity measurements.
\newline
We perform a dedicated analysis to study XY non-factorization on proton-proton data collected in 2022 at $\sqrt{s} = 13.6$~TeV by the CMS experiment~\cite{PAS}. A detailed examination of the shape of the bunch convolution function is presented, studying various biases, and choosing the best-fit analytic 2D functions to finally obtain the correction and its uncertainty.

\end{abstract}

\ccode{PACS numbers:} 07.05.Kf, 29.20.-c, 29.20.D-, 29.27.-a, 29.27.Bd, 29.27.Fh, 29.90.+r

\section{Introduction}

\subsection{What is luminosity?}
The instantaneous luminosity ($\mathcal{L}_\mathrm{inst}$) is an important quantity in experimental particle physics. It is proportional to the frequency of physical processes, therefore it can be measured as the fraction of the observed rate of some quantity ($R$) and a corresponding detector-dependent (calibration) parameter, the so-called visible cross-section ($\sigma_\mathrm{vis}$). On the other hand, luminosity can be expressed using the colliding beam parameters:
\begin{equation}
	\mathcal{L}_\mathrm{inst} = \frac{R}{\sigma_\mathrm{vis}} = \frac{f N_1 N_2}{2\pi \Sigma_\mathrm{x} \Sigma_\mathrm{y}}
	\label{Linst},
\end{equation} 
with the machine revolution frequency $f\approx 11245$~Hz, the bunch populations ($N_{1,2}$), and the horizontal and vertical beam overlap sizes ($\Sigma_\mathrm{x,y}$). These widths are measured using the so-called van der Meer (vdM) scans~\cite{vdM}.

\subsection{About vdM scans in a nutshell}
\begin{figure}[!h]
	\centerline{
	\includegraphics[width=0.33\textwidth]{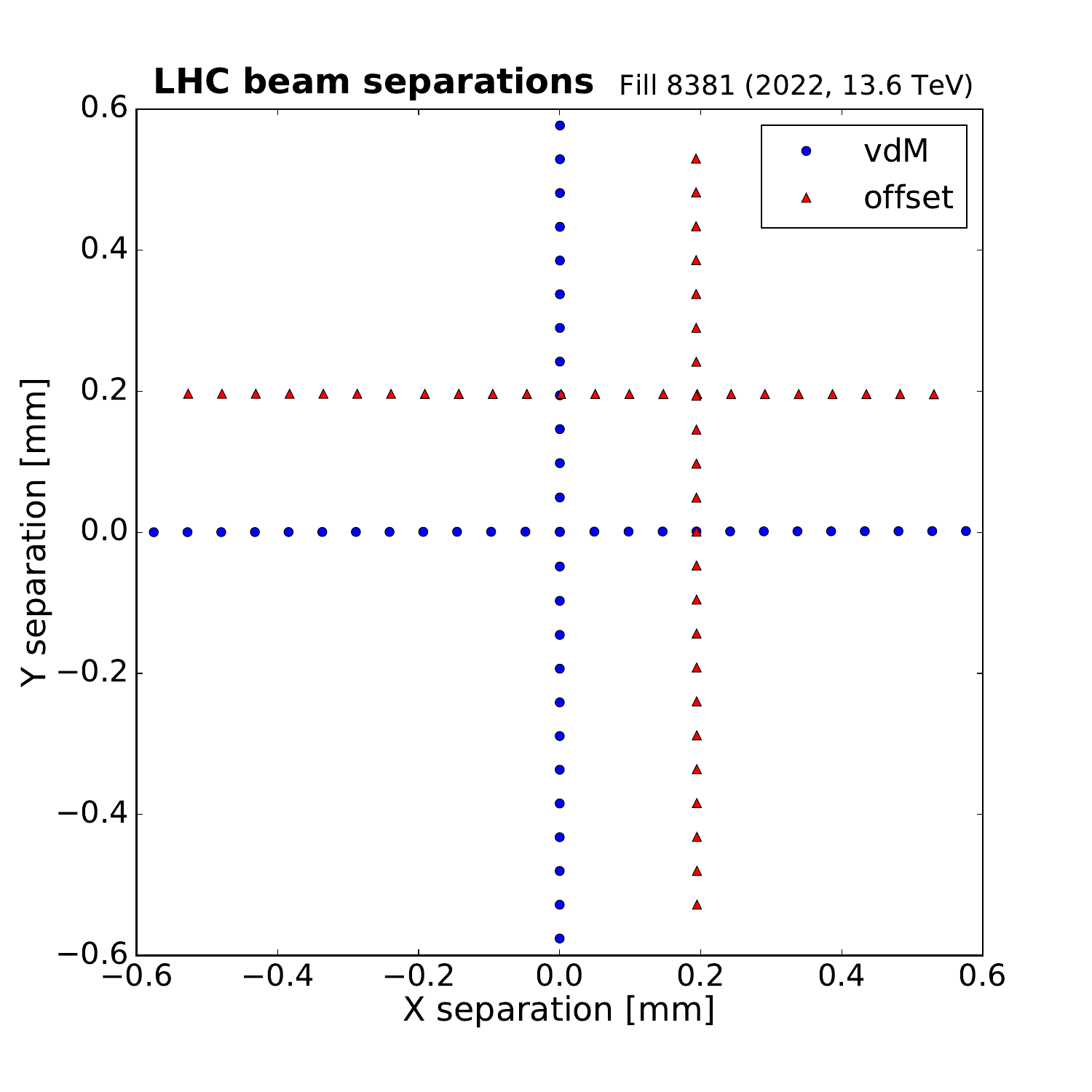}
        \includegraphics[width=0.33\textwidth]{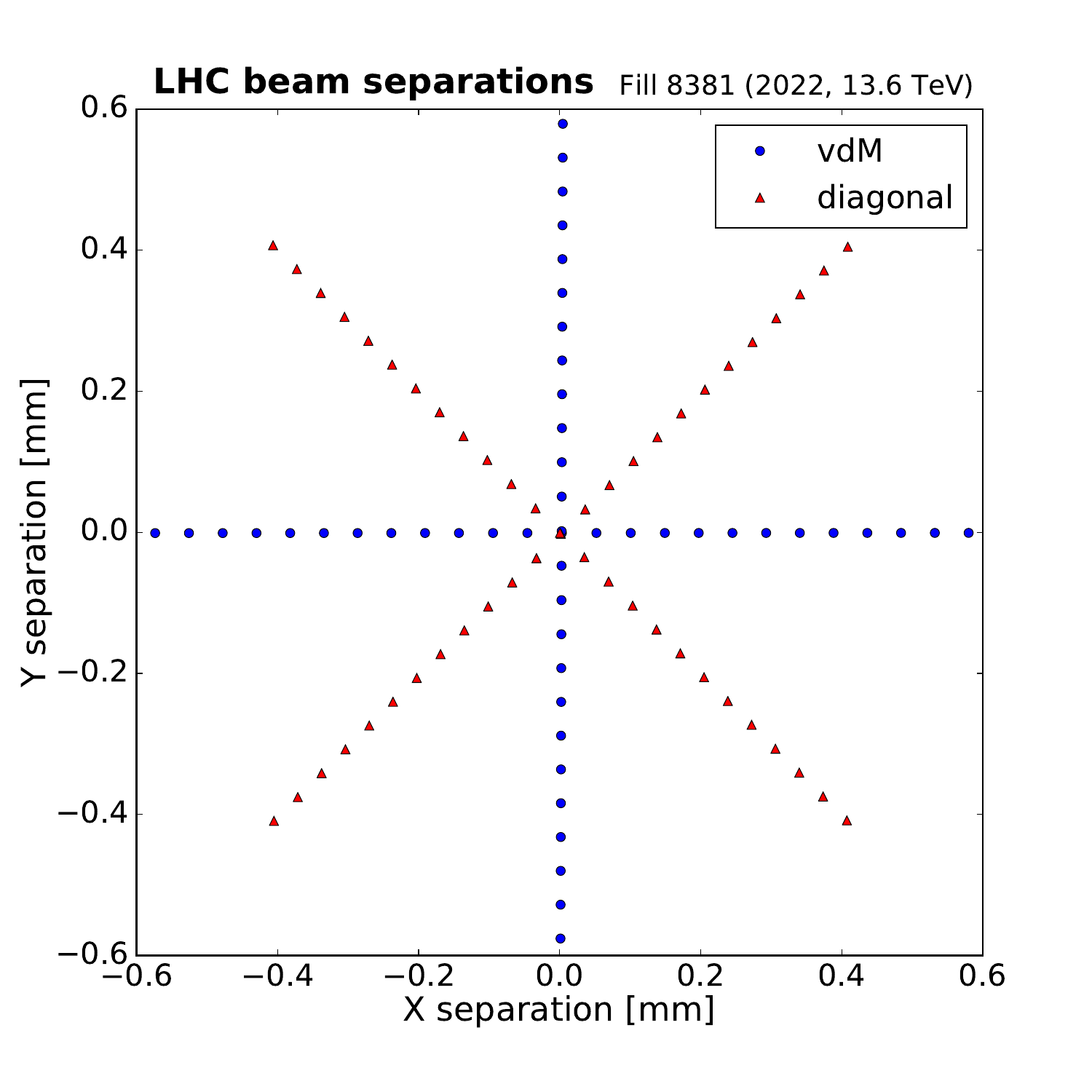}
        \includegraphics[width=0.33\textwidth]{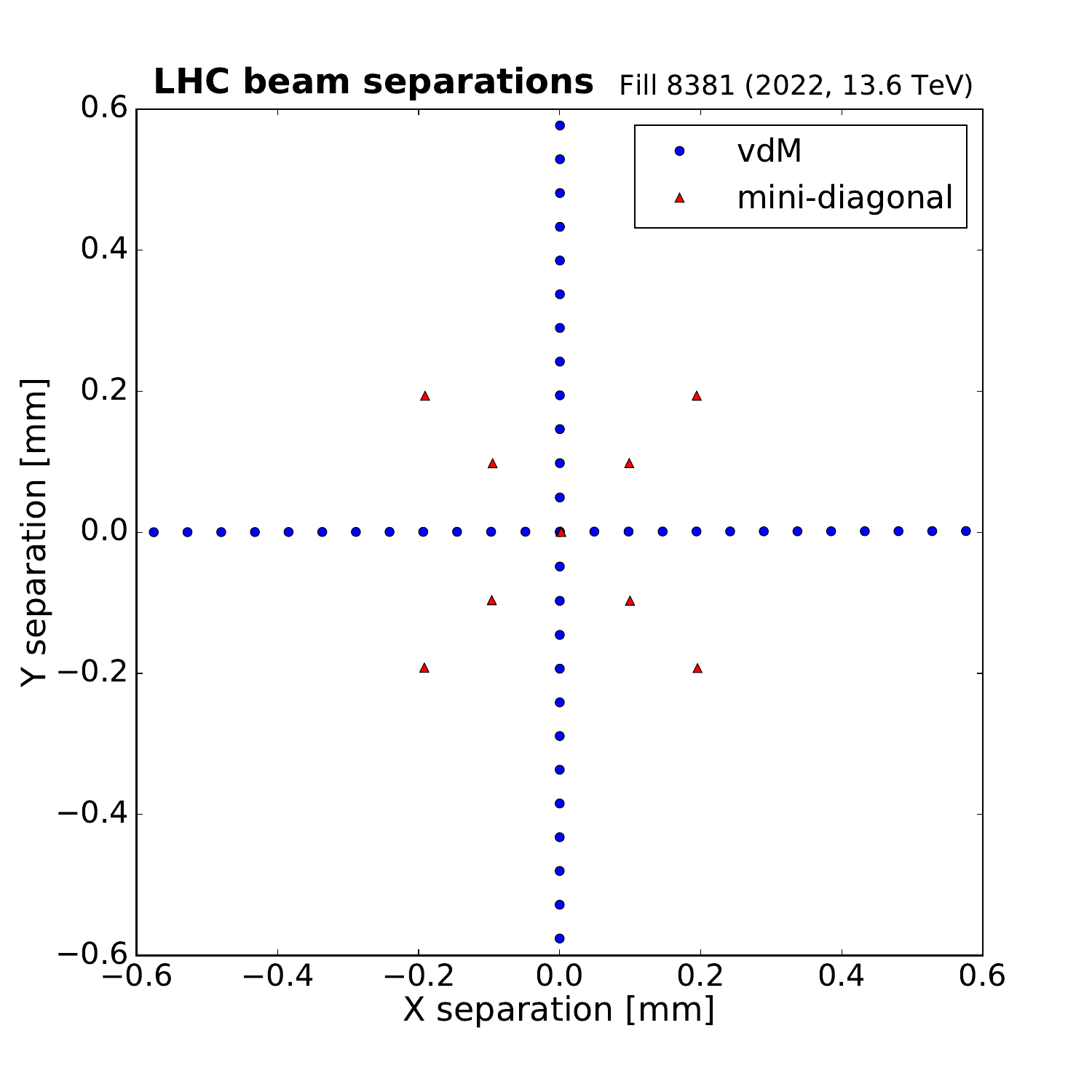}}
	\caption{Nominal transverse beam separations (defined by the currents of the LHC dipole corrector magnets) in $X$ and $Y$ directions during scans to illustrate the different scan types performed in 2022. The separation points (the distances by which the beams are separated) for vdM scans in $X$ and $Y$ directions are shown in blue, while for the off-axis scan points are marked in red. The latter ones belong to (in order from left to right) offset, diagonal and mini-diagonal scans. \label{OffsTogether}}
\end{figure}

At the Large Hadron Collider (LHC) two bunches (from oppositely circulating beams) collide. In vdM scans these bunches are separated with well-defined distances within the transverse $(x,y)$ plane, resulting in different rates measured by a detector, depending on the amount of overlap between the bunches (ie.\ how head-on the collisions are). Changing these transverse separation distances the vdM scans provide a rate distribution along the axes as
\begin{equation}
	R(0,y)\quad\mathrm{and}\quad R(x,0).
	\label{eq:vdM}
\end{equation}
The nominal separation values for the on-axis vdM scans are shown in Fig.~\ref{OffsTogether}, together with that of other, special off-axis scans (more details in the next section).

Then the on-axis vdM data is separately fitted in the two directions with some variations of the Gaussian function, from which the best fitting one is chosen and used ultimately for each scan. From these fits a width parameter ($\Sigma_\mathrm{x,y}$) is obtained, from which the $\sigma_\mathrm{vis}$ can be calculated as
\begin{equation}
	\sigma_\mathrm{vis} = \int\int R(x,y) \mathrm{d}x\mathrm{d}y \sim 2\pi\cdot \frac{\max_\mathrm{x} R(x,0) + \max_\mathrm{y} R(0,y)}{2} \cdot\Sigma_\mathrm{x}\Sigma_\mathrm{y}.
	\label{SigVis}
\end{equation}

\subsection{The XY Factorisation Bias}
In Eq.~\ref{SigVis}, however, the approximation is exact only if the convolution shape is factorizable, meaning that
\begin{equation}
	R(x,y) = f_1(x)\cdot f_2(y)
	\label{Fact}
\end{equation}
stands. This assumption was shown to be violated in previous CMS measurements, see e.g. Ref.~\cite{vdM} and necessitates the correction of the emerging XY factorization bias.

\begin{figure}[!h]
        \centerline{
        \includegraphics[width=0.45\textwidth]{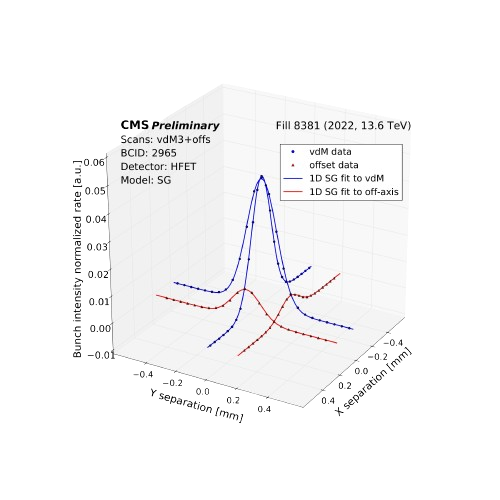}
        \hspace{-1.3cm}\includegraphics[width=0.45\textwidth]{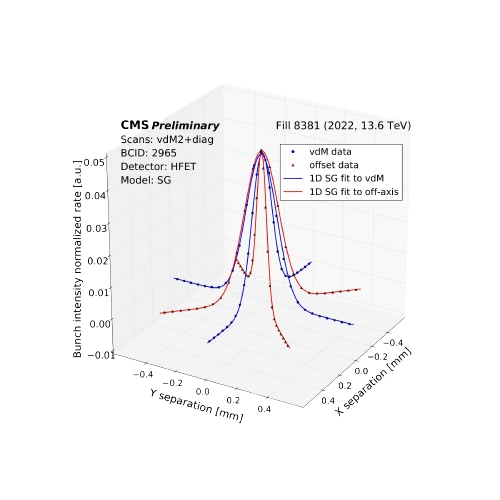}
        \hspace{-1.3cm}\includegraphics[width=0.45\textwidth]{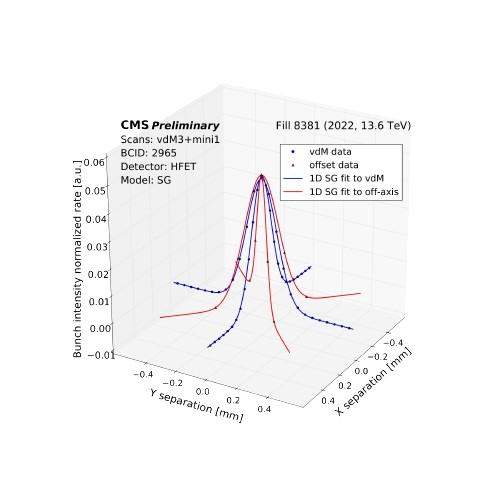}}
	\caption{Visualization of the HFET data at the scan points connected by 1D single Gaussian fits in on-axis vdM (blue) and off-axis (red) scan pairs. A simultaneous 2D fit is performed in each case to the luminometer rates to study the shape and the factorization property of the beam overlap area.
    \label{3D}}
\end{figure}

One possibility to measure this bias quantitatively is to combine the on-axis with off-axis scan data to obtain a more complete grid (as it is shown in Fig.~\ref{OffsTogether}) mapping the two-dimensional (2D) surface of the measured rates (shown in Fig.~\ref{3D} respectively to Fig.~\ref{OffsTogether}).
These points then can be fitted in 2D and the value for the bias can be obtained from a simulation based on the fitted shapes.

\section{Analysis}

The analysis workflow in details is presented through the example of the 2022 proton-proton dataset collected at 13.6~TeV collision energy~\cite{PAS, DPS}.

\subsection{Input data}
\begin{figure}[!h]
        \centering
        \includegraphics[width=\textwidth]{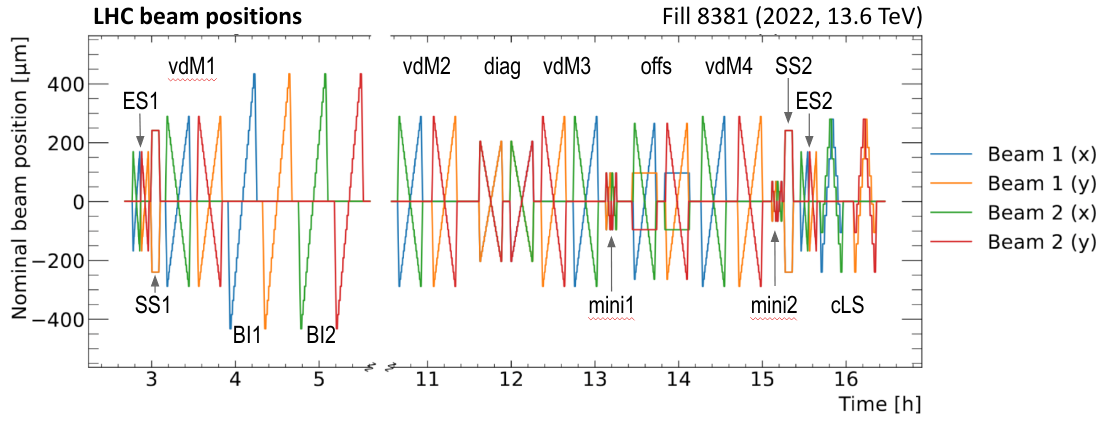}
	\caption{Nominal transverse beam positions (defined by the currents of the LHC dipole corrector magnets) as a function of time during LHC fill 8381. The labels of the various scan pairs are also given. Each scan pair contains two scans orthogonal to each other (e.g.\ horizontal vdM2X and vertical vdM2Y). Abbreviations: short on-axis vdM-like "emittance" scan (ES), super-separation with no overlap between the beams (SS), on-axis van der Meer scan (vdM), beam imaging scan with a stationary and a scanning beam (BI), off-axis diagonal scan (diag), mini-diagonal scan (mini), offset scan (offs), constant length scale scan with the two beams moving in the same direction (cLS). When several scans of the same type are performed, a number also follows the abbreviations to create unique labels. During the time gap between 5:36 and 10:36 CMS was taking head-on data. \label{vdMScans}}
\end{figure}

Following the aforementioned method based on off-axis scans to sample the tails of the beam overlap are for obtaining a quantitative measure for the factorization bias, the analysis uses the following scan data (referring to labels shown in Fig.~\ref{vdMScans}).  All off-axis scans are exploited, including the offset~\cite{offs}, the diagonal~\cite{diag}, and both mini-diagonal scans. Each of them is paired with one or two close-by on-axis vdM scans from 
vdM2-vdM4 to minimize the effect due to the changing beam conditions with time.

The input data already contains various corrections obtained from the CMS BRIL~\cite{Run3det} vdM framework (FW) dedicated to the analysis of beam scan data, in particular, the background contribution to the rates, e.g., from detector noise; the contribution from so-called ghost and satellite~\cite{vdM} charges to the bunch intensities; the beam orbit based on measurements by the LHC beam position monitors; and the transverse position length scales are taken into account. 
The effect of the electromagnetic interactions between the bunches~\cite{PAS, BBPaper}  was estimated to be small compared to other sources of bias and is not considered further.

\subsection{Analysis flow}
The analysis flow contains many steps and throughout cross-checks at each one.

As the first step, \textbf{1D fits} (regarding only one transverse scan direction at a time) are performed, similarly to the FW, but \textbf{also evaluated for the off-axis scans}.

With a method called \textbf{rate matching (RM)}, which compares the rates of the offset scans to the vdM ones at their intersection points, one aims to verify or obtain a more precise \textbf{orbit drift (OD) correction} (wrt.\ that of the FW) to have the best possible alignment between the on- and off-axis scan pairs. Calculating the values with which one shall shift the offset scan beam separations in the scanning direction (around 3~$\mu$m in the 2022 analysis with respect to the linear OD corrected values), one also gets a measure for the possible uncertainty in both the scanning and non-scanning directions that is used later to study systematics by varying, in the case of the diagonal scan data, the nominal (or, in the case of the offset scan data, the corrected "central") separation values with this amount in each direction, giving nine alignment options. In addition, for the offset scan data, we also use the uncorrected nominal alignment, resulting in total ten variations.

\begin{figure}[!h]
        \centering
        \includegraphics[width=0.66\textwidth]{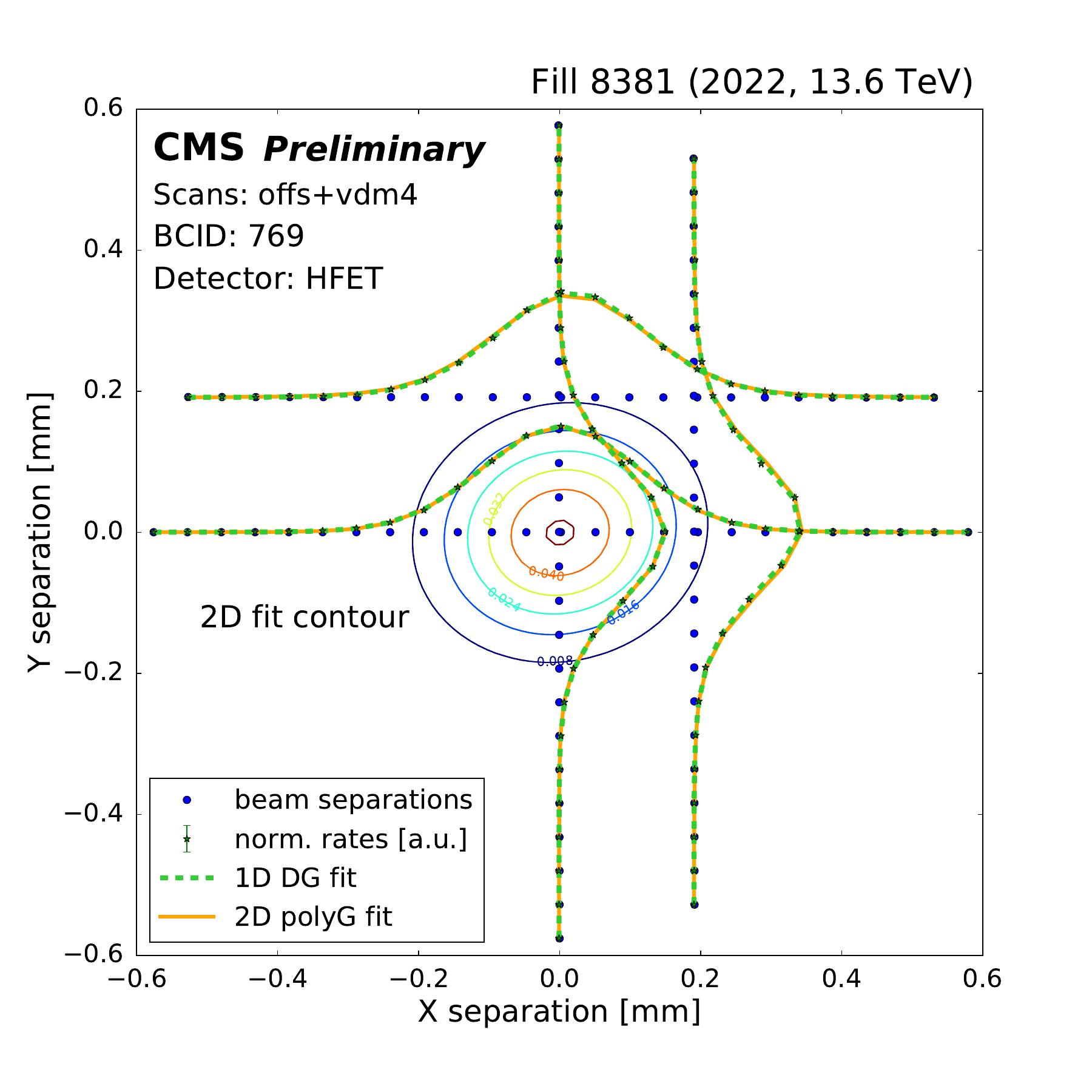}
	\caption{Comparison of the 1D ($X$ or $Y$) vdM fit results using double Gaussian functions (dashed green lines) with the 2D ($X - Y$) fit result using a polynomial Gaussian shape (rainbow contours are shown to illustrate the shape and orange lines for the slices). The scan points in the $X$ separation - $Y$ separation plane are indicated (blue dots). The HFET data is shown in a vdM scan and offset scan pair. The rates are normalized with different factors for better visibility for the vdM and offset scan slices. \label{Slices}}
\end{figure}

Next, the evaluation of the \textbf{2D fits} (Fig.~\ref{Slices}) follows, where various shapes have to be considered, for instance \textsl{single Gaussian} (SG, \ref{app2}), \textsl{super Gaussian} (supG, \ref{app3}), \textsl{double Gaussian} (DG, with various constraints, \ref{app10}), \textsl{q-Gaussian} (qG, \ref{app7}) or \textsl{polynomial Gaussian} (polyG, \ref{app5}). Regarding the DG shape we consider multiple different cases: fitting with the linear correlation $\varrho=0$ requirement (DG\_RhoFixed); and a two-step fitting starting with the DG\_RhoFixed results and refitting by released $\varrho$ (DG\_RhoReleased). We also consider various reparametrizations aiming to improve the fit stability. These are DG\_TiltFixed (\ref{app11}), where we fix X/Y width parameter ratios of the 2D Gaussians to the same value, and we also fix their linear correlation coefficients to be equal ($\mathrm{d}\varrho=0$). The other version is DG\_AxisFixed (\ref{app13}); fixing the angle between the axes of the DG functions to zero (i.e.\ requiring they have the same orientation).

The heart of the analysis is the determination of the bias using  simulated vdM scans. Since the 2D fit function parameters with their covariance matrix from the luminometer data (being either vdM+diag, vdM+mini or vdM+offs) are known at this stage, one can sample the parameter space to get representative 2D distributions for the beam overlap area of the vdM “data”. Varying the fitted parameters according to their uncertainty leads to a random 2D shape. Now the question naturally arises: How well would a vdM scan perform on such a shape? To answer this, one can calculate the rate at pre-defined (the nominal) separation points using this randomized 2D shape (with data-based uncertainty on it), then do a "hypothetical" vdM fit (in 1D) on this artificial data. Comparing the $\sigma_\mathrm{vis}$ from the simulated vdM with the exact $\sigma_\mathrm{vis}$ of the randomized 2D shape will give the factorisation correction as their ratio (minus one). Meanwhile, the standard deviation over many randomized shapes will lead to the fit uncertainty. This method thus also lets the uncertainty of the 2D fit propagate onto the correction in the end.

Finally, the whole method is validated with a closure test.

\subsection{Results}

CMS has several independently calibrated luminometers which can be all used for the analysis in best case scenario. Since they are supposed to measure the same quantity (luminosity), their prediction is expected to correlate bunch crossing by bunch crossing. It is very unlikely for instance that for one of the luminometers a (significantly) positive correction is obtained, while for the very same bunch crossing (BCID) on the basis of another luminometer the correction is negative. Therefore, checking the correlation between detectors provides a validation of the results and can signal problems with the performance of a particular detector if the results obtained by it do not correlate with the others, (while the rest show a decent correlation between each other).

\begin{figure}[!h]
	\centering
        \includegraphics[width=\textwidth]{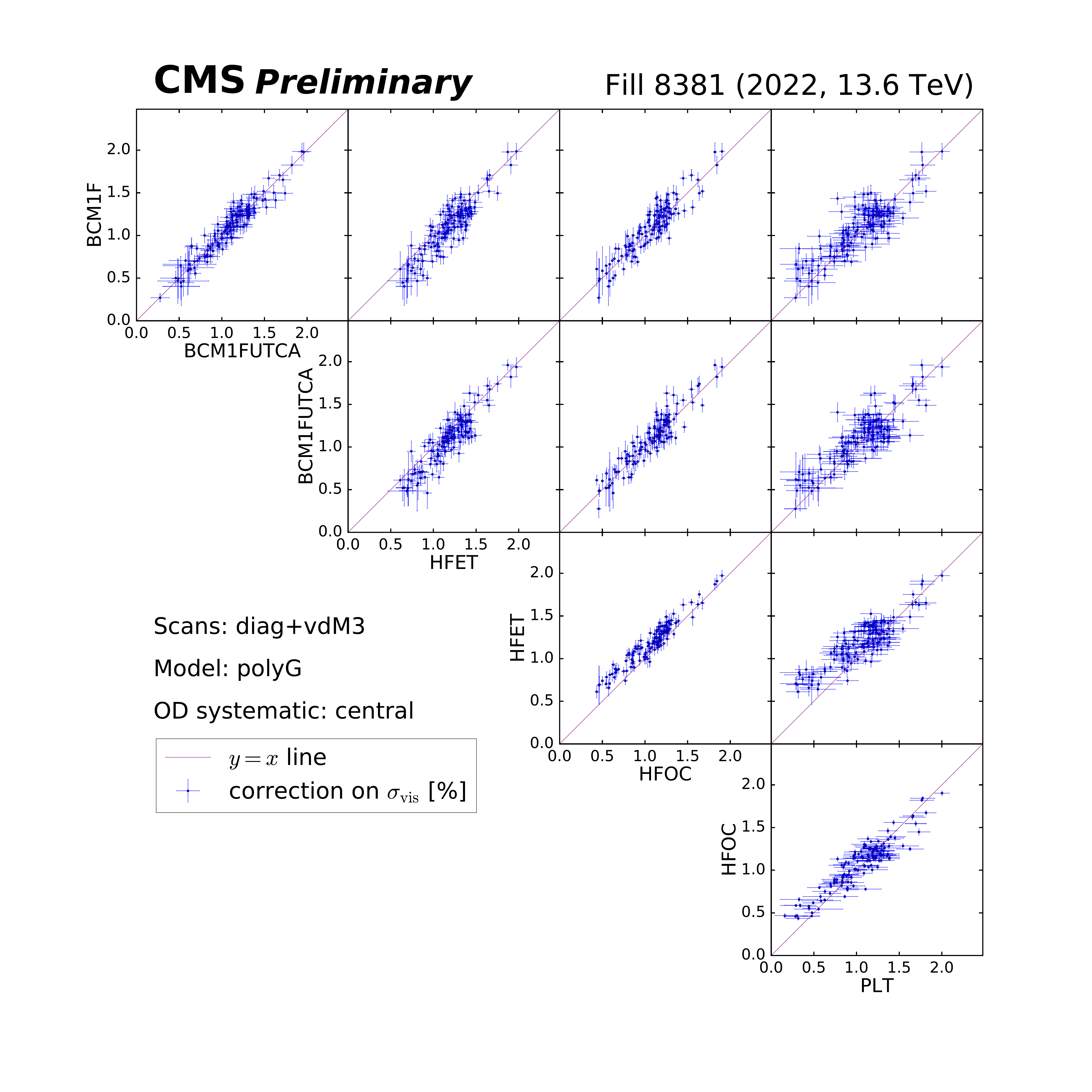}
	\caption{Correlation between the factorization corrections on the visible cross sections measured using different luminometers. As an example, the results for the polynomial Gaussian model are shown using the data from the diagonal scan and the closest (third) vdM scan following it, with the central values for the orbit drift (OD) corrected separations. Each subplot contains the measurements for all colliding bunch pairs with good fit results. The red lines show the diagonal with equal corrections to guide the eye. The uncertainties are statistical only. \label{CorrDets}}
\end{figure}

In our example, fortunately, a strong correlation can be seen for all stable shape models for all scan-pair pairs (for an example see Fig.~\ref{CorrDets}, more details on model stability later). All stable measurements are used to define the final correction and its uncertainty, which must be independent of the luminometer as it can only depend on the bunch shapes.

\begin{figure}[!h]
	\centering
        \includegraphics[width=0.75\textwidth]{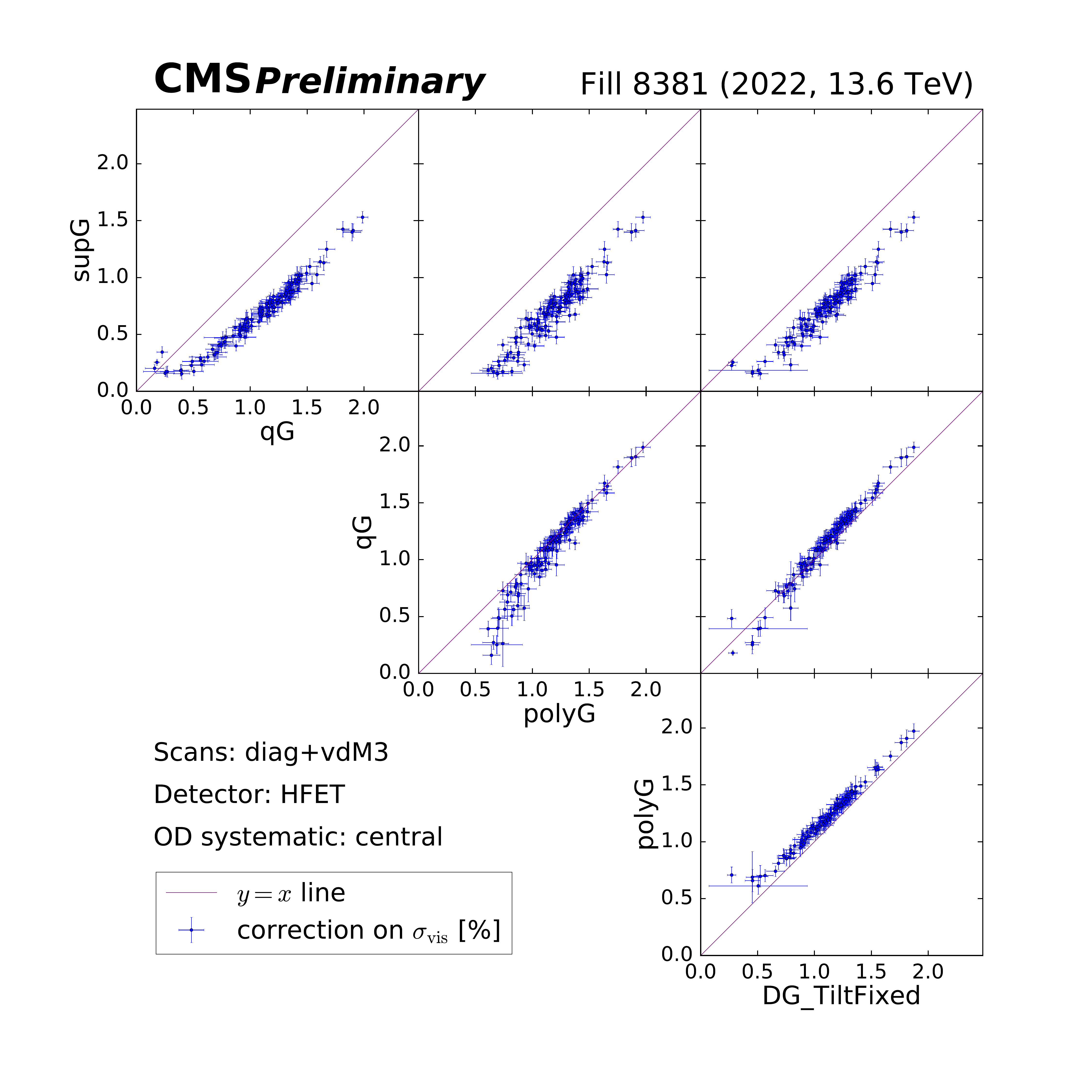}
	\caption{Correlation between the factorization corrections on the visible cross sections measured using different 2D shape models. As an example, the results are shown for the HFET data from the diagonal scan and the closest vdM scan following it, with the central values for the orbit drift (OD) corrected separations. Each subplot contains the measurements for all colliding bunch pairs with good fit results. The red lines show the diagonal with equal corrections to guide the eye. The uncertainties are statistical only. \label{CorrModels}}
\end{figure}

A very similar cross-check is performed for the fitted shapes. Checking their behaviour helps to decide which ones can be considered to obtain the final correction value. Having a look at the models (Fig.~\ref{CorrModels}) one can see strong correlations, even though the supG model gives consistently smaller values. The four models included in the figure are used to define the final correction and its uncertainty in the end.

The aforementioned models are used in the end, however, others were also studied. The SG has too few parameters to describe well the 2D shape, therefore it does not make a part of the final set of models. Regarding several of the DG models, they showed an instability, oscillating between two minimas. This motivated their exclusion from the final set of models to derive the central value of the correction.

\begin{figure}[!h]
	\centerline{
        \includegraphics[width=0.5\textwidth]{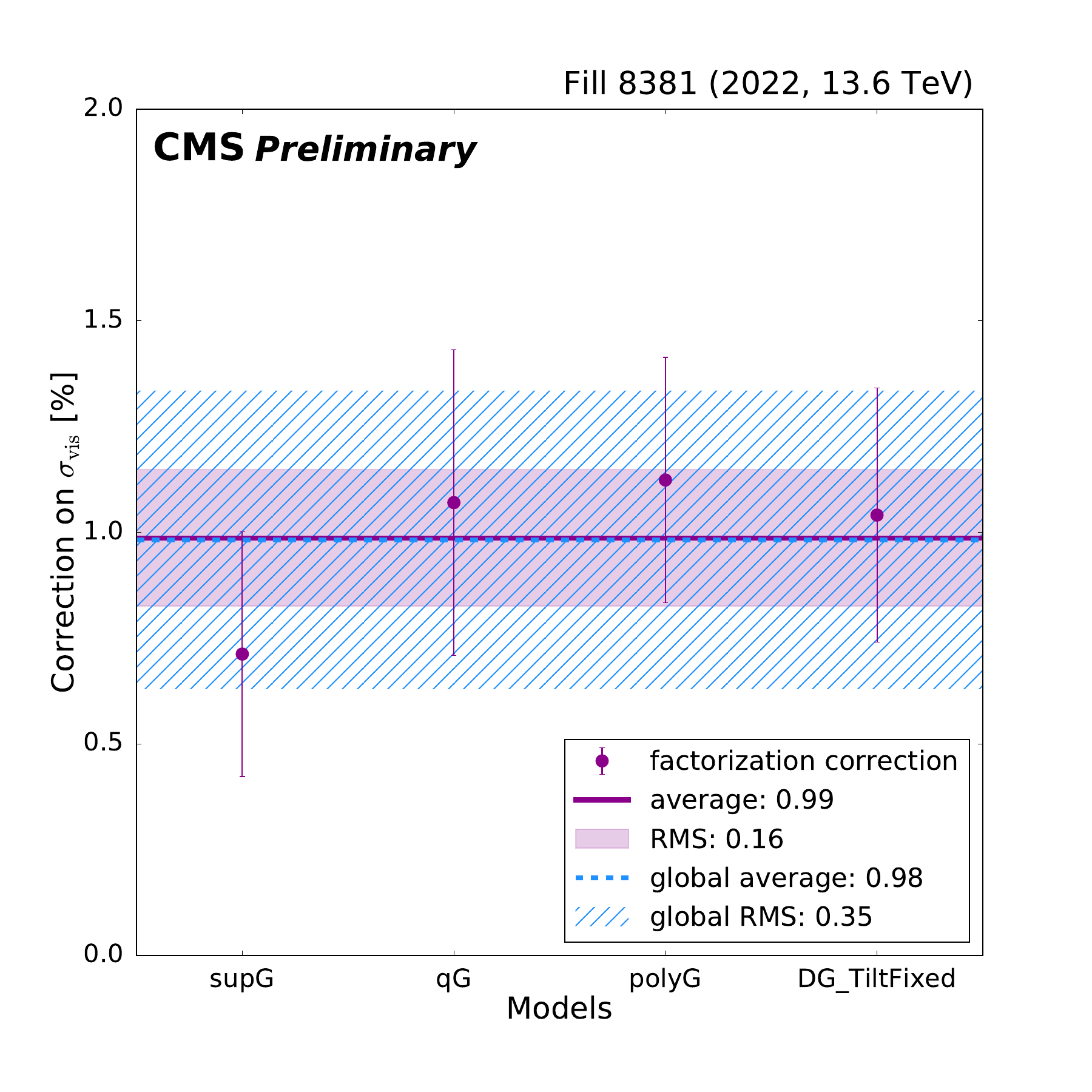}
        \includegraphics[width=0.5\textwidth]{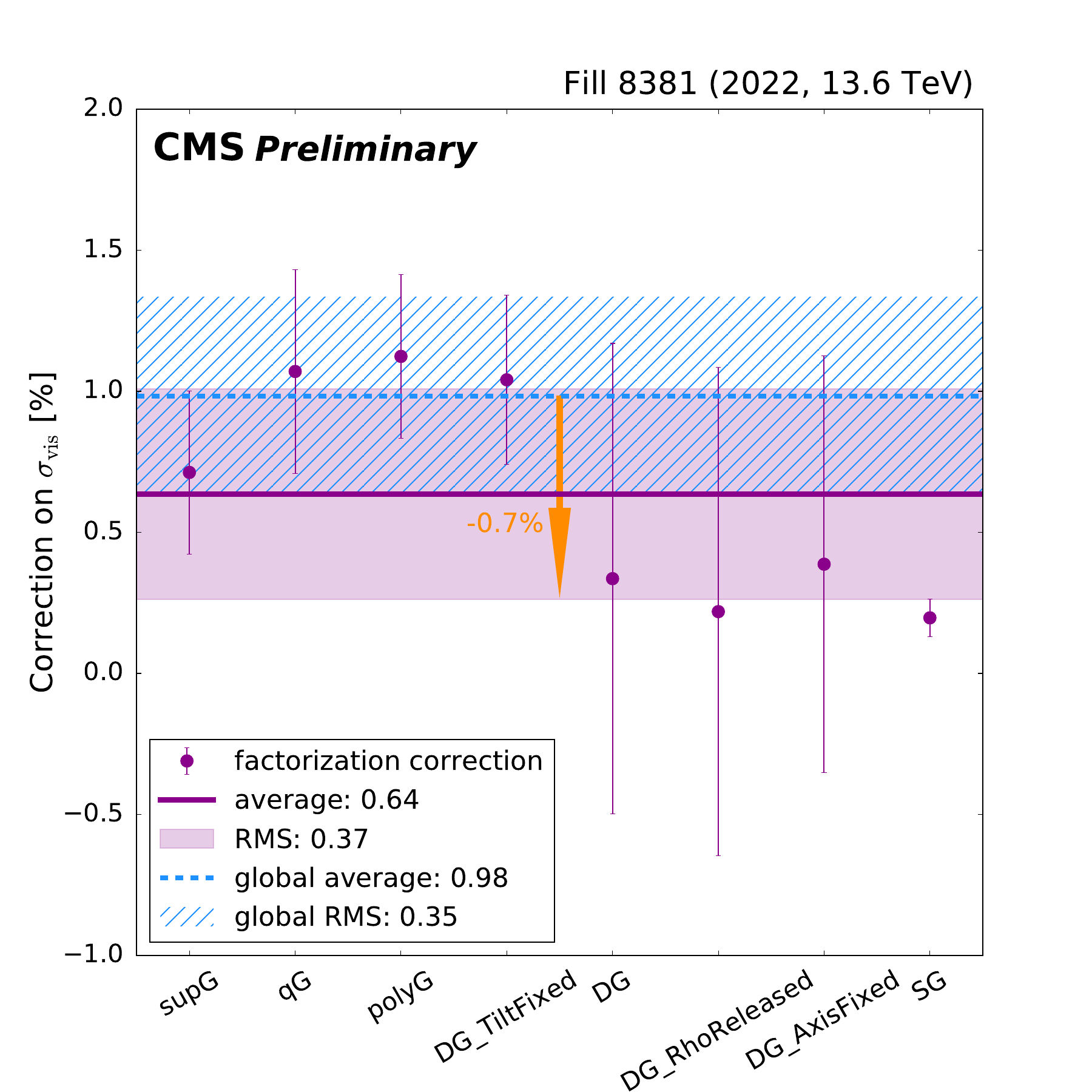}}
	\caption{Factorization corrections on the visible cross section $\sigma_\mathrm{vis}$ for all the various 2D models tested. Only the stable models are used for the central value of the final correction (marked as “global average” with its RMS as uncertainty in the legend). An additional model systematic uncertainty of 0.7\% accounts for the results of the three unstable double Gaussian models (see on the right). The single Gaussian model cannot account for the dominant source of non-factorization thus is not included in the global average. Each value is an average over the measurements performed by the five luminometers, using the data of seven on-axis - off-axis scan-pair pairs, and 144 colliding bunch pairs, with nine (or ten) orbit drift configurations - in the case of the offset scan data, one can look at the rate matching corrected alignment of the scan pairs, giving an extra degree of freedom. The average over the models with its RMS as uncertainty is given in the legend, together with the global average and its RMS over more than 350000 measurement variants. \label{VarModels}}
\end{figure}

However, they were used to define an uncertainty related to the model choices as illustrated by Fig.~\ref{VarModels} (right).  
We take the difference between the average of the results using the stable models and the average from the DG unstable models as an additional model systematics of 0.7\%. This is the dominant uncertainty which accounts for the possible bias from restricting the models to those that have elliptic contours (and give stable fits), and suppressing those that could give cross-like contours (and show instability). Fig.~\ref{VarModels} also illustrates that the model dependence for the four stable 2D models has a sub-dominant contribution to the observed variations.

\begin{figure}[!h]
	\centering
        \includegraphics[width=0.66\textwidth]{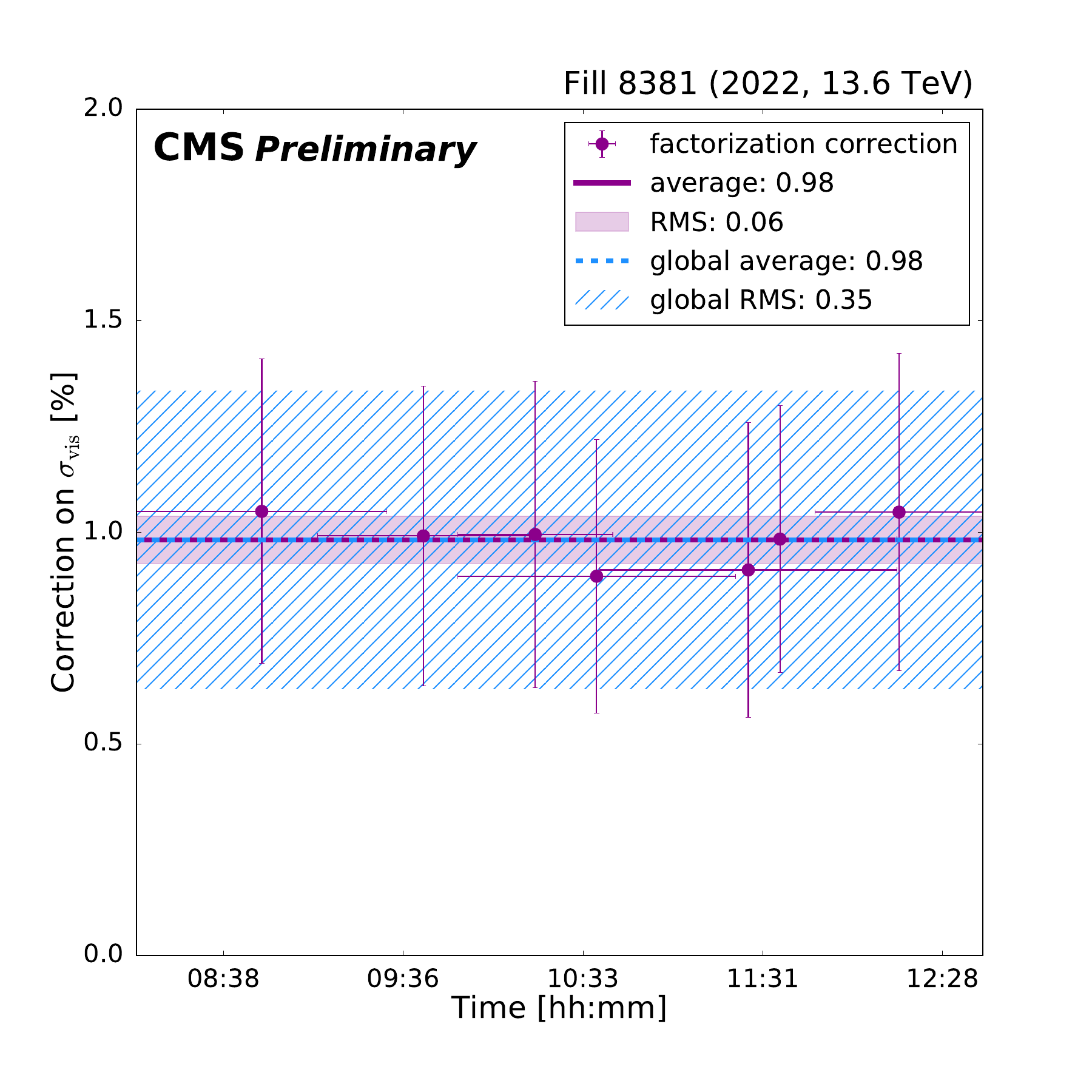}
	\caption{Factorization corrections on the visible cross section for the various on-axis- off-axis scan-pair pairs arranged by the data taking time. Each value is an average over the measurements performed by the five luminometers on the data of 144 colliding bunch pairs using nine (or ten) orbit drift configurations - in the case of the offset scan data, one can look at the rate matching corrected alignment of the scan pairs, giving an extra degree of freedom -, and four stable 2D shape models. The points from left to right correspond to vdM2+diag, diag+vdM3, vdM3+mini1, vdM3+offs, mini1+vdM4, offs+vdM4 and vdM4+mini2 combinations. The average over the scan-pair  combinations with its RMS as uncertainty is given in the legend, together with the global average and its RMS over more than 350000 measurement variants. \label{TimeDep}}
\end{figure}

Another important task is to check for possible time dependence, because it can happen that the bunch shape, thus the correction value as well, changes as the measurement proceeds. During these measurements, however, negligible (if any) time-dependence can be seen (Fig.~\ref{TimeDep}).

\begin{figure}[!h]
	\centering
        \includegraphics[width=0.66\textwidth]{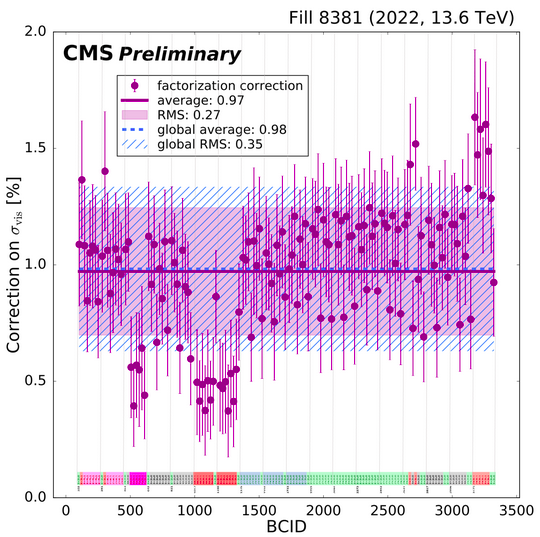}
	\caption{Factorization corrections on the visible cross section for the various bunch crossing identification (BCID) values arranged from 0 to 3563 along the LHC orbit. Each value is an average over the measurements performed by the five luminometers, using the data of seven on-axis - off-axis scan-pair pairs, with nine (or ten) orbit drift configurations - in the case of the offset scan data, one can look at the rate matching corrected alignment of the scan pairs, giving an extra degree of freedom -, and four stable 2D shape models. The average over the BCIDs with its RMS as uncertainty is given in the legend, together with the global average and its RMS over more than 350000 measurement variants. \label{BCIDDep}}
\end{figure}

One can also take a look at how the correction looks like BCID-by-BCID. In Fig.~\ref{BCIDDep} it can be read off that the BCID dependence gives the vast majority of the RMS. The well-visible pattern corresponds to the collision scheme in LHC (i.e.\ in which experiments the participating bunches collide) which is indicated by the color code just above the horizontal axis. Every fourth BCID tend to have a lower correction, which is likely related to the filling sequence which has a periodicity of four in the PS Booster, due to its four rings.

\section{Conclusion}

The final result is (0.98~$\pm$~0.35~$\pm$~0.7)\%, or rounded to (1.0~$\pm$~0.8)\%. Here the global RMS of 0.35\% comes from all 144~$\times$~4~$\times$~(9~or~10)~$\times$~7~$\times$~5 measurements, where in case of the offset scan data, one can look at the rate matching corrected alignment of the scan pairs, giving an extra degree of freedom and the dependence on the BCID is considered to have a physics motivation, since it correlates with the collision pattern at the LHC. The extra uncertainty of 0.7\% is also taken into account to cover the lack of knowledge of the true shape (while the model uncertainty in the global RMS is coming only from the four well-behaving models). The closure test gives a negligible contribution, below 0.1\%. The uncertainties in more detail are summarized in Tab.~\ref{Uncerts}.
\begin{table}[ph]
	\tbl{Uncertainties on the final result (third column, in percentage). Their sources are listed in the first column together with the number of the variant in the second column. There were 144 bunch crossings, seven different scan-pair pairs, nine (or ten) different alignments of the scan pairs to check OD systematics (in the case of the offset scan data, one can look at the rate matching corrected alignment, giving an extra degree of freedom), five luminometers and four models chosen for the final result. We considered an extra uncertainty obtained from three additional models to cover the lack of knowledge of the true shape. From the closure test the contribution is negligible. }
	{\begin{tabular}{@{}lcc@{}} \toprule
	Source of uncertainty & Number of variants    & Uncertainty [\%] \\
	\colrule
	BCID                  & 144                   & 0.27 \\
	Time (scan-pair pairing)   & 7                     & 0.05 \\
    OD systematics        & 10 offset, 9 diagonal & 0.05 \\
    Luminometer           & 5                     & 0.02 \\
	Model (well-behaving) & 4                     & 0.16 \\
   \colrule
	Model (additional)    & 3                     & 0.70 \\
   \colrule
    Closure test           &                      & $<$0.1 \\
	\botrule
    Total           &                      & 0.98 \\
	\botrule
	\end{tabular} \label{Uncerts}}
\end{table}

\section*{Acknowledgments}

This work was partially supported by the National Research, Development and Innovation Office – NKFIH, under contract numbers K 143460, K 146913, and K 146914.

\clearpage
\appendix

\section{Formulas}

To shorten the formulas describing our 2D fit functions we introduce the term
\begin{eqnarray}
        r^2 =  \frac{\Big( \frac{x-\mu_\mathrm{x}}{\sigma_\mathrm{x}} \Big)^2 + \Big( \frac{y-\mu_\mathrm{y}}{\sigma_\mathrm{y}} \Big)^2 - 2\varrho \frac{x-\mu_\mathrm{x}}{\sigma_\mathrm{x}}\frac{y-\mu_\mathrm{y}}{\sigma_\mathrm{y}} }{ 1 - \varrho^2 }.
\end{eqnarray}

\subsection*{Single Gaussian (SG) including a correlation term}
The parameters of the function are written out as arguments to simplify the formulas below.
\begin{eqnarray}
    SG_\mathrm{2D}\big(x,y; V, \mu_\mathrm{x}, \mu_\mathrm{y}, \sigma_\mathrm{x}, \sigma_\mathrm{y}, \varrho\big) = \frac{V}{2\pi \sigma_\mathrm{x} \sigma_\mathrm{y}\sqrt{1-\varrho^2}}
                                                               \exp\Big(-\frac{r^2}{2}\Big).
    \label{app2}
\end{eqnarray}

\subsection*{Super Gaussian (supG)}
\begin{eqnarray}
        supG_\mathrm{2D}\big(x,y; V, \mu_\mathrm{x}, \mu_\mathrm{y}, \sigma_\mathrm{x}, \sigma_\mathrm{y}, \varrho, p\big) = \frac{V}{N}\mathrm{exp}\Big( -\frac{r^{2p}}{2}\Big),
        \label{app3}
\end{eqnarray}
where 
\begin{eqnarray}
        N &= 2\pi \sigma_\mathrm{x} \sigma_\mathrm{y} \sqrt{1-\varrho^2} \cdot \frac{\Gamma(1/p)}{p}2^{1/p}.
\end{eqnarray}

\subsection*{Polynomial Gaussian (polyG)}
Here the constant amplitude of the Gaussian is replaced by a symmetric fourth order polynomial:
\begin{eqnarray}
    polyG_\mathrm{2D}\big(x,y; V, \mu_\mathrm{x}, \mu_\mathrm{y}, \sigma_\mathrm{x}, \sigma_\mathrm{y}, \varrho, a_2, \bar{a}_4\big) = \frac{V}{\tilde{N}} (1 + a_2 r^2 + a_4 r^4)
    \mathrm{exp}\Big( -\frac{r^2}{2} \Big)
    \label{app5}
\end{eqnarray}
where
\begin{eqnarray}
    \begin{array}{rcl}
        \tilde{N} =  2\pi \sigma_\mathrm{x} \sigma_\mathrm{y} \sqrt{1 - \varrho^2} \cdot (1 + 2 a_2 + 8 a_4)\,, \\[8pt]
        a_4 =
        \begin{cases}
            \bar{a}_4 & \quad \textrm{if } a_2>0 \\
            \bar{a}_4 + a_2^2/4   & \quad \textrm{otherwise}.
        \end{cases}
    \end{array}
\end{eqnarray}
For positivity $\bar{a}_4>0$ is required.

\subsection*{Q-Gaussian (qG)}
\begin{eqnarray}
    qG_\mathrm{2D}(x,y) = \frac{V}{C(q) \cdot \sigma_\mathrm{x} \sigma_\mathrm{y} \sqrt{1 - \varrho^2}} \cdot e_\mathrm{q}\Big( -\frac{r^2}{(4-2q)}\Big),
    \label{app7}
\end{eqnarray}
where
\begin{eqnarray}
    e_\mathrm{q}(x) = \big[1 + (1-q)x\big]_+^\frac{1}{1-q},
\end{eqnarray}
and  $[x]_+$ is equal to $x$ if $x>0$, and 0 otherwise. Furthermore
\begin{eqnarray}
    C(q) =
    \begin{cases}
            \frac{4 - 2q}{1 - q}\pi\frac{\Gamma\big(\frac{2-q}{1-q}\big)}{\Gamma\big(\frac{2-q}{1-q}+1\big)} & q < 1 \\
            \frac{4 - 2q}{q - 1}\pi\frac{\Gamma\big(\frac{1}{q-1}-1\big)}{\Gamma\big(\frac{1}{q-1}\big)} & \textrm{otherwise}.
    \end{cases}
\end{eqnarray}
To handle the poles near $q=1$ a linear interpolation is used in the $|q-1|<0.006$ range. A smaller range can be used by utilizing the $\ln(\Gamma(x))$ ("gammaln") function~\cite{Ghoshdastidar:2014}.

\subsection*{Double Gaussian (DG)}
This is the sum of two SGs, with the two means fixed to be the same value:
\begin{eqnarray}
    \begin{split}
        DG_\mathrm{2D}(x,y) = V \cdot \Bigg(&v_\mathrm{R} \cdot SG_\mathrm{2D}\Big(x,y; 1, \mu_\mathrm{x}, \mu_\mathrm{y}, \sigma_\mathrm{x}^{(1)}, \sigma_\mathrm{y}^{(1)}, \varrho^{(1)}\Big) + \\        
                          &(1-v_\mathrm{R}) \cdot SG_\mathrm{2D}\Big(x,y; 1, \mu_\mathrm{x}, \mu_\mathrm{y}, \sigma_\mathrm{x}^{(2)}, \sigma_\mathrm{y}^{(2)}, \varrho^{(2)}\Big)\Bigg),
    \end{split}
    \label{app10}
\end{eqnarray}
where $v_\mathrm{R}$ denotes the volume ratio.

\subsection*{Double Gaussian for fixed tilt (DG\_TiltFixed)}
A reparametrization of the DG function, where requiring $\varrho^{(1)}=\varrho^{(2)}$ is possible:
\begin{eqnarray}
    \begin{split}
        DG_\mathrm{2D}^{TiltFixed}(x,y) = V \cdot \Bigg(&v_\mathrm{R} \cdot SG_\mathrm{2D}\Big(x,y; 1, \mu_\mathrm{x}, \mu_\mathrm{y}, \sigma_\mathrm{x}^{(1)}, \sigma_\mathrm{y}^{(1)}, \varrho+\mathrm{d}\varrho\Big) + \\        
                      &(1-v_\mathrm{R}) 
                      SG_\mathrm{2D}\Big(x,y; 1, \mu_\mathrm{x}, \mu_\mathrm{y}, \sigma_\mathrm{x}^{(2)}, \sigma_\mathrm{y}^{(2)}, \varrho-\mathrm{d}\varrho\Big)\Bigg),
    \end{split}
    \label{app11}
\end{eqnarray}
where
\begin{eqnarray}
    \begin{aligned}
    \sigma_\mathrm{x}^{(1)} &= \sigma_\mathrm{x} \sqrt{s_\mathrm{R}}\sqrt{s_\mathrm{A}},\\
    \sigma_\mathrm{y}^{(1)} &= \sigma_\mathrm{y} \sqrt{s_\mathrm{R}}/\sqrt{s_\mathrm{A}},\\
    \sigma_\mathrm{x}^{(2)} &= \sigma_\mathrm{x} /\sqrt{s_\mathrm{R}}/\sqrt{s_\mathrm{A}},\\
    \sigma_\mathrm{y}^{(2)} &= \sigma_\mathrm{y} /\sqrt{s_\mathrm{R}}\sqrt{s_\mathrm{A}}.
    \end{aligned}
\end{eqnarray}
$v_\mathrm{R}$ refers to the volume ratio, $s_\mathrm{R}$ to the sigma ratio, $s_\mathrm{A}$ to the sigma asymmetry that controls the "crossedness" of the two components.

\subsection*{Double Gaussian for fixed axis (DG\_AxisFixed)}
An angular reparametrization of the DG function, where requiring the rotation of the two components to be identical is possible:
\begin{eqnarray}
    \begin{split}
        DG_\mathrm{2D}^{AxisFixed}(x,y) = V \cdot \Bigg(&v_\mathrm{R} \cdot SG_\mathrm{2D}^{(\alpha)}\Big(x,y; 1, \mu_\mathrm{x}, \mu_\mathrm{y}, \sigma_\mathrm{x}^{(1)}, \sigma_\mathrm{y}^{(1)}, \alpha+\mathrm{d}\alpha\Big) + \\        
                      &(1-v_\mathrm{R}) 
                      SG_\mathrm{2D}^{(\alpha)}\Big(x,y; 1, \mu_\mathrm{x}, \mu_\mathrm{y}, \sigma_\mathrm{x}^{(2)}, \sigma_\mathrm{y}^{(2)}, \alpha-\mathrm{d}\alpha\Big)\Bigg),
    \end{split}
    \label{app13}
\end{eqnarray}
where $SG_\mathrm{2D}^{(\alpha)}$ refers to the following parametrization:
\begin{eqnarray}
    SG_\mathrm{2D}^{(\alpha)}\big(x,y; V, \mu_\mathrm{x}, \mu_\mathrm{y}, \sigma_\mathrm{x}, \sigma_\mathrm{y}, \alpha\big) = \frac{V}{2\pi \sigma_\mathrm{x} \sigma_\mathrm{y}\sqrt{1-\varrho^2}} \cdot
                                                               \exp\Big(-\frac{r^2_\alpha}{2}\Big),
\end{eqnarray}
where 
\begin{eqnarray}
    r^2_\alpha = 
\begin{pmatrix}
  x - \mu_\mathrm{x}\\ 
  y - \mu_\mathrm{y}
\end{pmatrix}^T
R(-\alpha)
\begin{pmatrix}
  \sigma_\mathrm{x}^{-2} & 0 \\ 
  0 & \sigma_\mathrm{y}^{-2}
\end{pmatrix}
R(\alpha)
\begin{pmatrix}
  x - \mu_\mathrm{x}\\ 
  y - \mu_\mathrm{y}
\end{pmatrix},
\end{eqnarray}
and $R(\alpha)$ is a $2\times 2$ rotation matrix.

\end{document}